\documentclass[aps,nofootinbib,twocolumn,showpacs,10pt]{revtex4}
\usepackage{amsmath}
\usepackage{graphicx}
\usepackage{psfrag}

\begin{document}
\title{Primordial galactic magnetic fields: An application of QCD
  domain walls.} 
\author{Michael McNeil Forbes}
\author{Ariel R. Zhitnitsky}
\affiliation{Department of Physics and Astronomy, University of
  British Columbia\\
  Vancouver, British Columbia, Canada, V6T 1Z1.}  
\begin{abstract}
  We present a mechanism for generating primordial magnetic fields
  with large correlation lengths on the order of 100 kpc today.  The
  mechanism is based on recently conjectured QCD domain walls or
  similar CP violating domain walls with QCD scale structure.  Such
  domain walls align the electric and magnetic dipole moments of the
  nucleons shortly after the QCD phase transition, producing
  electromagnetic fields correlated along the domain walls. Through
  the Kibble mechanism, the domain walls attain Hubble-scale
  correlations which they transfer to the aligned electromagnetic
  fields.  Due to the CP violation, the Hubble-scale walls produce
  helical (non-zero Chern-Simons) magnetohydrodynamic turbulence which
  then undergoes an inverse cascade, allowing the correlation lengths
  to grow to $100$ kpc today.  We present an estimate the magnitude of
  the generated electromagnetic fields in terms of the QCD parameters.
  We also discuss some other unexplained astrophysical phenomena which
  may be related to this mechanism. In particular, we discuss the
  relation between primordial magnetic fields and the
  Greisen-Zatsepin-Kuzmin (GZK) cutoff violations. We also outline
  how, by creating inhomogeneities in the nucleon density, QCD domain
  walls may lead to inhomogeneous big bang nucleosynthesis (IBBN)
  explaining the $\Omega_B$ excess recently measured through cosmic
  microwave background (CMB) distortions.
\end{abstract}

\maketitle

\section{Introduction}
The purpose of this paper is to present a careful argument of the
mechanism outlined in our letter \cite{Forbes:2000gr} to generate
large scale cosmic magnetic fields.  It is an extended and updated
version of the letter \cite{Forbes:2000gr} and the conference
proceedings \cite{Forbes:2000ew} and should be considered to be the
definitive version of these papers.

Many different observations suggest that there exist substantial
(microgauss) magnetic fields in the universe today
\cite{Kronberg:1994vk,Beck:1996zs}, however, there has yet to emerge a
theory which adequately explains the origins of these fields.  Most of
the data on large scale astrophysical magnetic fields comes from the
observation of synchrotron radiation emitted in galaxies.  This
radiation is plane polarized, and as it passes through magnetic
fields, the plane of polarization rotates due to the Faraday effect:
an effect which depends on the frequency of the radiation and the
strength and orientation of the magnetic fields.  By comparing several
sources or radiation with different frequencies, one can extrapolate
to determine the original plane of polarization and then estimate the
magnetic field strengths.

What is striking, is not just the existence of magnetic fields, but
that they appear to be microgauss fields which have correlations as
large as $500$ kpc in clusters.  To put this into perspective, the
luminous core of galaxies have typical scales of up to $10$ kpc while
it is estimated that the galactic dark matter halos extend to $50$
kpc.  Thus, it seems that galactic dynamo mechanisms cannot produce
these large correlation lengths.

The current models for producing these fields involve two main
processes: 1) dynamical amplification and/or generation of magnetic
fields by galactic processes (galactic mechanisms) and 2) primordial
mechanisms which take place prior to gravitational structure
formation.  The galactic mechanisms are primarily based on
gravitational dynamos, although there are suggestions that supernovae
or other stellar phenomena may play a role.  While it is likely that
galactic dynamos amplify fields, it seems difficult to account for the
large scale correlations of the magnetic fields when only galactic
mechanisms are considered.  It is also not certain that galactic
mechanisms can generate magnetic fields: instead they serve only as an
amplifier requiring seed fields to be present for the dynamo to work.

The inadequacies of the galactic mechanisms have lead to many
proposals that the magnetic fields may have a primordial origin.  In
this case, some process in the early universe (typically at a cosmic
phase transition or during inflation) is thought to generate magnetic
turbulence.  This turbulence then sustains itself as the universe
expands and what we observe are the remnants of this turbulence.
Most primordial sources, however, also produce fields which end up
with very small correlations today or which are very weak.

Most likely, a complete picture of the history of astrophysical
magnetic fields requires some primordial inputs as well as
amplification through gravitational dynamics.  In this paper, however,
we discuss a primordial mechanism which seems to naturally produce
fields of $100$ kpc correlations today.  In combination with dynamic
amplification mechanisms, we hope that this mechanism might provide a
solid foundation for the theory of large-scale astrophysical magnetic
fields.

At this point we would like to refer the reader to the several reviews
and sample papers in this field.  The primary discussions of
observations, which contain reviews of the theory, are presented in
\cite{Kronberg:1994vk,Beck:1996zs}.  Good current reviews are given in
\cite{Grasso:2000wj,Tornkvist:2000ay,Olesen:1997jh}.  Many different types of
primordial mechanisms are discussed, for example: Inflationary
mechanisms \cite{Turner:1988bw}, cosmic strings
\cite{Vachaspati:1991tt,Gasperini:1995dh}, charge asymmetries
\cite{Dolgov:1993mu,Joyce:1997uy}, and phase transitions
\cite{Baym:1996fk,Sigl:1997dm}.  The evolution of primordial magnetic
fields is discussed in
\cite{Olesen:1997jh,Brandenburg:1996fc,Olesen:1997ts}.  In particular,
the inverse cascades discussed here will be important for our
mechanism.

\section{Overview of the Mechanism}
The mechanism that we propose has the following core components:
\begin{enumerate}
\item Sometime near the QCD phase transition, $T_{\mathrm{QCD}}\approx
  1$ GeV, domain walls form which can interact with QCD scale physics.
\item These domain walls rapidly coalesce until there remains, on
  average, one domain wall per Hubble volume with Hubble scale
  correlations.
\item Baryons interact with the domain walls and align their spins
  along the walls.  The Hubble scale correlation of the domain walls
  thus induces a Hubble scale correlation in the spin density.
\item The anomalous magnetic and induced electric dipole moments of
  the baryons generate helical electromagnetic fields also
  correlated on the Hubble scale.
\item The domain walls move rapidly and vibrate, effectively filling
  the Hubble volume with helical magnetic turbulence with a Hubble
  scale correlation.
\item The domain walls decay and the electric fields are screened
  leaving magnetic turbulence with Hubble scale correlations.
\item As the universe expands, an ``inverse cascade'' mechanism
  transfers energy from small to large scale modes, effectively
  increasing the resulting correlation lengths but diluting the field
  strengths.
\item Galactic dynamos amplify the fields in galaxies, but the fields
  should also persist in the extra-galactic media.
\end{enumerate}
The idea that domain walls might generate magnetic fields is not
original.  For example, 
it was suggested that standard axion domain walls could be
ferromagnetic in \cite{Iwazaki:1997bk}, however mechanism discussed
their seems to be flawed: The scale of the standard axion walls is of
the order $m_a^{-1}$ which is at least some twelve orders of magnitude
larger than the QCD scale $\Lambda_{\mathrm{QCD}}^{-1}$ ($m_a$ is the
axion mass).  It is hard to see how these walls can efficiently affect
QCD physics at the temperatures that were present in the early
universe where thermal fluctuations will destroy all
coherence.\footnote{For some other discussions about the magnetic
  properties of the domain walls, see \cite{Cea:1998tu}
  and reference therein. We should note, however, that in all these
  discussions, the most difficult problem of generating large scale
  correlations has not been addressed.}

Another problem with proposals including standard axion domain walls
is that these walls must decay to prevent cosmological problems
\cite{Vilenkin:1994paper}.  There are still questions about how the
standard so-called $N\neq 1$ axion domain walls can decay.

In this paper, we outline the properties that domain walls must have
to generate sufficient magnetic seed fields.  The exact source could
be one of several types of walls, including modified axion domain
walls, or entirely different types of walls.  To be concrete, we
present our model in terms of a recently conjectured quasi-stable QCD
domain wall \cite{Forbes:2000et} which may exist, either independently
of axion physics, or along with a dynamical axion adding additional
QCD scale structure to the standard axion domain walls.  These domain
walls are characterized by a transition in the singlet $\eta'$ field
which has a size and energy scale set by $\Lambda_{\rm QCD}$.  Hence,
QCD domain walls can directly couple to QCD physics.  In addition, the
singlet field transition at the center of the wall induces an
effective non-zero CP violating $\theta$ background which in turn will
induce an electric dipole moment and alter the magnetic dipole moment
in the fermions \cite{Crewther:1979pi} so that both the electric and
magnetic dipole moments of all the particles are on the same order. In
the presence of these anomalous dipole moments, the cancellations
discussed in \cite{Voloshin:1996jw,Voloshin:2000jp} in the domain wall
background are no longer a problem.

Another crucial aspect of our mechanism is the ``inverse cascade''
which governs the evolution of the magnetic fields after they are
formed.  This mechanism was suggested by Cornwall
\cite{Cornwall:1997ms}, discussed by Son \cite{Son:1998my} and
confirmed by Field and Carroll \cite{Field:1998hi}.  It is based on
the idea that magnetic helicity (Abelian Chern-Simons number)
$H=\int{\vec{\bf A}\cdot\vec{\bf B}}{\rm d}^3{x}$ is approximately
conserved in the universe where temperatures are higher than 
$T_0\approx 100$ eV.  This conservation of helicity causes energy to
cascade up the turbulent modes increasing the energy in large scale
modes and increasing the effective correlation length of the
turbulence.  The importance of helicity was originally demonstrated by
Pouquet and collaborators \cite{Pouquet:1976}.  Without this helical
inverse cascade, there is no known way to generate large correlations
fields today from sub-Hubble scale fields in the early phase transition
and one must consider super-Hubble scale correlations resulting form
inflationary schemes.  It turns out, however, that Hubble scale
correlations at the QCD phase transition (the last major phase
transition) provide correlations on the order of $100$ kpc today: thus
it is natural to consider QCD physics as the source of primordial
fields (earlier physics can only produce even smaller correlations).

Perhaps the least understood aspect of this mechanism concerns the
dynamics of the domain walls and the interactions of the domain walls,
nucleons and electric and magnetic fields.  As we shall show, all of
these components interact on the same scale of $\Lambda_{\rm QCD}$ and
hence there are complication back-reactions and nonlinear dynamics.  We
presently do not have the tools to fully analyze these features, but
we present here in detail quantitative estimates and calculations
which we believe are good estimates of the scale of the effects.

The result is a mechanism which naturally produces magnetic fields
today with $l\sim 100$ kpc correlations and with strengths of 
\begin{equation}
  \label{eq:*}
  B_{\mathrm{rms}}\sim
  \frac{10^{-9}{\rm  G}}{\xi\Lambda_{\rm QCD}} ,~~~~ l\sim 100~{\rm kpc}
\end{equation}
where the parameter $1\leq\xi\Lambda_{\rm QCD}\ll10^{19}$ depends on
the dynamics of of the domain walls as discussed in
section~(\ref{sec:generation}).  If the correlation $\xi$ turns out to
be small, then this mechanism might generate detectable extra-galactic
fields, otherwise we still require a galactic dynamo to amplify the
fields.  In any case, however, these seeds still maintain the large
scale correlations of the observed fields, and it seems that even if
$\xi$ is large, the resulting fields may be strong enough to seed the
galactic dynamos \cite{Zeldovich:1980}.

We shall begin by discussing the inverse cascade mechanism in
section~(\ref{sec:evolution}) and then give estimates of the field
strengths in an idealized case of static, flat walls.  Finally, we
shall discuss the dynamics of the domain walls and describe the whole
process, justifying the mechanism.

\section{Evolution of Magnetic Fields}
\label{sec:evolution}
Given a stable magnetic field configuration in the universe, one might
na{\"\i}vely expect the size of the correlations of the field to expand
with space as governed by the universe's scale parameter $l\propto
R(T)$ and the field strength to be correspondingly diluted
$B\propto R(T)^{-2}$.  It was discovered by Pouquet and collaborators
\cite{Pouquet:1976}, however, that if the magnetic fields have a
non-zero helicity (Abelian Chern-Simons number) $H=\int{\vec{\bf
    A}\cdot\vec{\bf B}}{\rm d}^3{x}$, then the fields will scale
differently.  Cornwall \cite{Cornwall:1997ms} suggested that helical
fields might undergo an inverse cascade.  The magnetohydrodynamic
(MHD) equations were subsequently studied by Son \cite{Son:1998my} who
derived the scaling relations (\ref{eq:scale}) presented below.  These
have subsequently been confirmed by Field and Carroll
\cite{Field:1998hi}.

The reason for the scaling is that the magnetic helicity $H$ is an
approximately conserved quantity in the early universe.  It is also
known that the small scale turbulent modes decay more rapidly than the
higher scale modes.  In order to conserve the helicity as the small
scale modes decay, the helicity must be transfered to larger modes and
with this transfer of helicity is a transfer of energy.  This is the
source of the inverse cascade.

To understand the origins of the conservation of helicity, note that
it is a topological quantity that describes the Gaussian linking
number of the vector potential lines of flux (see for example
\cite{Choudhuri:1998}).  In a perfectly conducting medium, these
lines of flux cannot cross, and hence there is no way to unlink the
flux lines: helicity is perfectly conserved.  Even when the
conductivity is finite, the helicity is also well conserved.

There is a direct analogy with fluid mechanics.  The equivalent there
is vorticity $\zeta=\int{\vec{\bf v}\cdot(\vec{\bf \nabla} \times
\vec{\bf v})}\mathrm{d}^3{x}$ which is the Gaussian linking number of the
fluid flow lines.  If the fluid has no viscosity, then $\zeta$ is
perfectly conserved.  Even viscous fluids, however, approximately
conserve vorticity.  This is why, for example, smoke rings and tornados
are so stable.

We shall not derive the scaling relationships here, instead we refer
the reader to \cite{Son:1998my,Field:1998hi}, however, we summarize here
the results:
\begin{subequations}
  \label{eq:scale}
  \begin{eqnarray}
    \label{eq:scaleb}
    B_{\mathrm{rms}}(T_{\mathrm{now}})&=&
    \left(\frac{T_0}{T_{\mathrm{now}}}\right)^{-2}
    \left(\frac{T_i}{T_0}\right)^{-7/3}B_{\mathrm{rms}}(T_i)
    \label{eq:brms1}\\
    \label{eq:scalel}
    l(T_{\mathrm{now}})&=&\left(\frac{T_0}{T_{\mathrm{now}}}\right)
    \left(\frac{T_i}{T_0}\right)^{5/3}l(T_i).
    \label{eq:corr1}
  \end{eqnarray}
\end{subequations}
These relate the initial field strength $B_{\mathrm{rms}}(T_i)$ with
initial correlation $l(T_i)$ to the present fields today
($T_{\mathrm{now}}\approx2\times10^{-4}$ eV)
$B_{\mathrm{rms}}(T_{\mathrm{now}})$ with correlation
$l(T_{\mathrm{now}})$.  During the period when the universe supports
turbulence (as indicated by a large Reynolds number ${\rm Re}$), the
inverse cascade mechanism functions and have the scalings $B\propto
T^{7/3}$ and $l\propto T^{-5/3}$ as indicated by the second factors in
(\ref{eq:scale}).  In the early universe, $\mathrm{Re}$ is very large
and the turbulence is well supported.  As the universe cools,
eventually, for temperatures below some $T_0$, the turbulence is no
longer well supported.  Exactly what the effective temperature $T_0$
the turbulence ceases is not clear: Son points out that at $T_0\approx
100$ eV, the Reynolds number drops to unity and thus turbulence is not
well supported because of the viscosity of the plasma
\cite{Son:1998my}.  We take this as a conservative estimate.  Field
and Carroll argue that the turbulence is force-free and so unaffected
by the viscosity.  Thus they take $T_0\sim 1$ eV at the epoch when the
matter and radiation energy densities are in equilibrium and argue
that the cascade may even continue into the matter dominated phase of
the universe.  If this is true, then it might be possible to increase
the correlation lengths of the fields by one or two orders of
magnitude from the conservative estimate (\ref{eq:*}).  In any case,
for temperatures lower than $T_0$, the turbulence and inverse cascade
are not supported and so we assume that the fields are ``frozen in''
and experience only the na\"{\i}ve scaling $l\propto T^{-1}$ and
$B\propto T^2$ indicated by the first factors in (\ref{eq:scale}).

As pointed out by Son \cite{Son:1998my}, the only way to generate
turbulence is either through a phase transition $T_i$ or through
gravitational instabilities.  Thus, until gravitational dynamos are
active, the scalings (\ref{eq:scale}) should be valid.  In any case,
galactic dynamos will amplify the fields, but will not affect the
correlation length, so in particular, (\ref{eq:scalel}) should be a
good estimate, regardless of galactic dynamics (the uncertainty come
during the transition period $T\sim T_0$ when the scaling laws
change. The estimate (\ref{eq:scalel}) is generally considered
conservative in this sense as the cascade likely continues for some
time past $T_0$ \cite{Field:1998hi}.

Now we consider the source of the magnetic turbulence at a phase
transition.  As we shall show, our mechanism generates Hubble size
correlations $l_i$ at a phase transition $T_i$.  In the radiation
dominated epoch, the Hubble size scales as $T_i^{-2}$.  Combining this
with (\ref{eq:corr1}), we see that $l_{\mathrm{now}}\propto
T_i^{-1/3}$; thus, the earlier the phase transition, the smaller the
resultant correlations.

The last phase transition is the QCD transition,
$T_i=T_{\mathrm{QCD}}\approx 0.2$ GeV with Hubble size
$l(T_{\mathrm{QCD}})\approx 30$ km.  With our estimates (\ref{eq:EB})
of the initial magnetic field strength $B_{\mathrm{rms}}(T_i)\approx
e\Lambda_{QCD}^2/(\xi\Lambda_{\rm QCD})\approx (10^{17}{\rm
  G})/(\xi\Lambda_{\rm QCD})$ we use Equations (\ref{eq:scale}) to
arrive at the estimate (\ref{eq:*}).  The meaning of the correlation
length $\xi$ will be discussed in detail later in
section~(\ref{sec:gen}).  The most important result here is that, as
long as one has a mechanism to generate Hubble scale correlations and
a maximally helical magnetic field at the QCD phase transition,
magnetic turbulence of $100$ kpc correlations is naturally produced.
The questions: `How can helical magnetic fields with Hubble-scale
correlations be produced at the QCD phase transition' and, `Are these
fields strong enough to account for the observed microgauss fields?'
will be addressed in the rest of this paper.  The estimate
(\ref{eq:*}) suggests, however, that even in the worst case of almost
maximal suppression $\xi\Lambda_{\mathrm{QCD}}\sim 10^{19}$, an
efficient galaxy dynamo may be able to amplify the fields to the
microgauss level.  In the best case, the mechanism would produce
measurable extra-galactic fields.

In either case, the important result is the generation of the $100$
kpc correlations: if observations show that the fields have much larger
correlations, then the proposed mechanism can only be salvaged if a
more efficient ``inverse cascade'' mechanism is shown to work between
$T_{\mathrm{QCD}}$ and now.  Having said this, one might consider the
electroweak or earlier phase transitions.  As we mentioned, the
earlier the phase transition, the smaller the resulting correlations
$l_{\mathrm{now}}\propto T_i^{-1/3}$.  For the electroweak transition,
the scaling (\ref{eq:scalel}) suggests that Hubble scale helical
fields could generate $100$ pc correlations today.  Thus it might be
possible that electroweak phenomena could act as the primordial
source, but this presupposes a mechanism for generating fields with
Hubble scale correlations.  Such a mechanism does not appear to be
possible in the Standard Model.  Instead, the fields produced are
correlated at the scale $T_i^{-1}$ which can produce only $\sim 1$ km
correlations today which are of little interest.

Thus, the previous analysis seems to suggest that, in order to obtain
magnetic fields with $100$ kpc correlation lengths, helical fields
must be generated with Hubble scale correlations near or slightly
after the QCD phase transition $T_{\mathrm{QCD}}$.  The same
conclusion regarding the relevance of the QCD scale for this problem
was also reached by Son, Field and Carroll
\cite{Son:1998my,Field:1998hi}.  Without further ado, we now present
our picture of the mechanism and justify the the estimate (\ref{eq:*}) of
the magnetic field strength.

\section{Generation of Magnetic Fields by Domain Walls}
\label{sec:generation}
The key players in this mechanism are domain walls which form shortly
after the QCD phase transition.  Details of the walls were presented
in \cite{Forbes:2000et} which will be summarized in
section~(\ref{sec:QCDdw}).  In sections (\ref{sec:align}) and
(\ref{sec:gen}) we shall show that these walls tend to align nuclear
magnetic and electric dipole moments along the plane of the wall.  An
important feature of the walls is that across the wall there is
maximal strong CP violation due to an induced nonzero $\theta$.
Because of this, the electric and magnetic dipole moments of the
nucleons are of the same order.  Thus, both neutrons and protons will
have non-zero electric and magnetic dipole moments and play a role in
generating the electromagnetic fields.

Because of the correlation between the electric and magnetic fields
along the domain wall, the generated fields have an induced helicity
as we shall examine in section~(\ref{sec:gen}).  This helicity has
the same sign along the entire domain wall and we expect that the
domain wall will fill the entire Hubble volume, thus the helicity will
be correlated on the Hubble scale.

Finally, the domain walls will decay as discussed in \cite{Forbes:2000et}
so that the universe is not dominated by domain walls today.  By this
point, however, the helical magnetic turbulence has been generated.

\subsection{Hubble Size Correlations}
\label{sec:dw}
The reason that we feel domain walls hold the key to explaining
primordial magnetic seeds is that in a short time they can generate
Hubble scale correlations.  The initial fields must have a Hubble
scale correlation or else there is no known way---even with the
inverse cascade---to generate the huge correlations today.  Let us
briefly summarize what we expect to be the behaviour of domain walls
at the QCD phase transition.  For a nice description of general domain
wall dynamics see \cite{Vilenkin:1994paper} from which most of these
results were derived.

\begin{enumerate}
\item Prior to the phase transition $T_i=\Lambda_{\mathrm{QCD}}$, the fields are
  in random fluctuations on the scale $T_i$ and domain walls are not
  present.
\item After the phase transition, however, the fields settle into
  their vacuum states.  Domains are formed where the fields are
  settling into different\footnote{In the case of QCD domain walls,
    the vacuum states are actually the same but the field
    configuration going from one domain to the next undergoes a
    classically stable transition.  This behaviour is qualitatively
    similar to the sine-Gordon model ${\mathcal{L}}_{\mathrm{SG}} =
    (\partial_\mu\phi)^2-\cos\phi$ where $\phi$ is interpreted as a
    phase so that the vacuum states $\phi = 2\pi n$ are actually
    identical.} vacuum states.  These domains are separated by domain
  walls and have a scale set by $\Lambda_{\mathrm{QCD}}$.
\item Numerical studies suggest that these small-scale domain walls
  rapidly merge increasing the correlation length of the walls.  This
  coarsening occurs simultaneously throughout space and the
  correlation length of the domain walls can increase faster than the
  speed of light.
\item The coarsening stops once the domain walls attain a Hubble
  scale.  On average, one ends up with one domain wall per Hubble
  volume, but which curls and moves through space, essentially filling
  the volume.
\end{enumerate}

It is these Hubble sized domain walls that can generate magnetic
turbulence with Hubble size correlations.  As we shall see below,
there are two types of domain walls corresponding to opposite field
transitions.  One we call a ``soliton'', and the other we call an
``anti-soliton''.  Together a soliton and an anti-soliton can annihilate,
but the coarsening essentially separates regions of solitons from
anti-soliton regions by a distance of the Hubble scale so that they do
not annihilate.  In section~(\ref{sec:helicity}) we shall show that the
solitons and anti-solitons are associated with helicity of the opposite
sign.  Thus, the domain walls effectively separate the helicity
generating a Hubble scale correlation length in the fields and in the
helicity.\footnote{There is some question about what conditions must be like at the phase
transition in order for domain walls to form.  QCD lattice simulations
suggest that at low densities (such as those present in the early
universe), the transition between the quark-gluon plasma and the
normal hadronic phase is a smooth crossover and that the critical
point sits at some finite density (see the recent review
\cite{Rajagopal:2000wf}).  If the rate at which the universe cools is
sufficiently slow, then it is possible that no domain walls will form.
In the preface to the paperback edition \cite{Vilenkin:1994paper},
the authors discuss this scenario as the Kibble-Zurek picture: to estimate
the size of the correlations produced, one must consider the
relaxation timescale $\tau(T)$ that it takes to establish correlations
on the scale $\xi(\tau)$.  The freezout temperature $T_f$ is
determined by the condition $\tau\sim t_D=|T-T_c|/|\dot{T}|$, i.e. when
the relaxation time is on the same order as the dynamical timescale of
temperature variations.  Since the true critical point is at a somewhat
higher densities than the universe, $t_D$ my be bounded from below and
if the cooling is sufficiently slow, it is possible that $\tau\ll t_D$
and domain walls will not form.  To estimate these effects requires a
better understanding of the dynamics of the domain walls and of the
phase transition than we presently have.  We assume that the dynamics
are such that domain walls do form and coalesce forming Hubble-scale
correlations as described in \cite{Vilenkin:1994paper}.}

\subsection{Essential Domain Wall Properties}
\label{sec:DWsources}
Thus, we can formulate a set of properties that must be satisfied by
domain walls if they are to be considered as sources for the primordial
magnetic fields described in this paper:
\begin{enumerate}
\item \label{enum:DWhubble} The walls must attain Hubble scale
  correlations near the QCD phase transition to generate the observed
  correlations.
\item \label{enum:DWscale} The walls must have structure on the scale
  of $\Lambda_{\mathrm{QCD}}$ in order to interact effectively with
  nucleons.
\item \label{enum:DWdiamagnetic} There must be some way to avoid the
  cancellations discussed in \cite{Voloshin:1996jw,Voloshin:2000jp} so
  that the walls are ferromagnetic rather than diamagnetic.
\item \label{enum:DWhelicity} The walls must somehow induce a definite
  helicity throughout the Hubble volume.
\item \label{enum:DWdecay} The domain walls must be unstable or have
  other features so that the problems of domain wall domination in the
  universe are avoided, but they must be stable enough that they can
  generate the appropriate fields.  They must also leave
  nucleosynthesis production ratios relatively unaffected.
\end{enumerate}

It seems to be a rather general property of cosmological domain wall
networks that they rapidly coarsen through the Kibble mechanism until
the walls have a Hubble-scale correlation length
\cite{Vilenkin:1994paper}.  Thus, criterion \ref{enum:DWhubble} should
be easily satisfied by almost all types of domain walls.  Criterion
\ref{enum:DWscale} rules out the standard axion domain walls discussed
in \cite{Iwazaki:1997bk,Huang:1985tt,Chang:1998tb}, however, there may
be features of these walls that have QCD scale which were previously
neglected.  In particular, Shifman and Gabadadze discuss a gluonic
core sandwiched at the center of axion domain walls
\cite{Gabadadze:2000vw} which they call D-walls.  This structure has a
QCD scale and may be able to align nuclear matter.

In our paper \cite{Forbes:2000et} we discuss another possibility: that
the $\eta'$ field may provide domain wall structure with QCD scale.
This structure, which we shall refer to as a QCD domain wall, can
exist within the standard axion domain wall providing the required
QCD scale structure, but can also exist, even if there is no axion
(unlike the D-walls of \cite{Gabadadze:2000vw} which require an axion
field).

A further property of QCD domain walls and axion domain walls is that,
at their center, there is strong CP violation.  This CP violation has
several effects as described in section~(\ref{sec:CP}).  In particular,
the CP violation induces an anomalous electric dipole moments in
nucleons.  Thus, CP violating domain walls can satisfy criterion
\ref{enum:DWdiamagnetic}.

Helicity is also associated with CP violation as it is a CP odd
quantity.  As discussed above, the soliton and anti-soliton domain
walls solutions have opposite CP.  Thus, each is
associated with opposite helicity.  Typically, the solitons and
anti-solitons separate spatially so that Hubble-sized regions are
filled with one type or another.  The helicity is generated through the
correlation of both electric and magnetic fields along the walls.
Thus, the Hubble scale spatial separation of soliton and anti-soliton
domain walls also separates the helicity and thus generates helical
turbulence with Hubble scale correlations satisfying criterion
\ref{enum:DWhelicity}.

Another major problem with axion domain walls is that most varieties
appear to be absolutely stable.  The $N=1$ axion model discussed
recently by Chang, Hagmann and Sikivie \cite{Chang:1998tb} has a decay
mode that satisfies the criterion \ref{enum:DWdecay}.  However, this
is not a phenomenologically acceptable model and, to date, the other
axion models are plagued by this problem.  If axion domain walls are
to be considered, then a satisfactory solution to this problem must be
found.

The QCD domain walls \cite{Forbes:2000et} without an axion, exhibits a
similar decay mode to the the $N=1$ axion model.  The scales, however,
are set by $\Lambda_{\mathrm{QCD}}$ rather than $m_a$ and so criterion
\ref{enum:DWdecay} must be address from the point of view: Do the
walls live long enough to generate the turbulence?  As addressed in
\cite{Forbes:2000et}, the answer may be yes.  The only decay mode is
through a nucleation process suppressed by quantum mechanical
tunneling.  Consequently, these walls may have a macroscopic lifetime
long enough to generate the fields.  In any case, however, they decay
fast enough to avoid affecting nucleosynthesis and other cosmological
effects.

Thus, there may be several types of domain walls that could act as
sources for primordial magnetic seed fields.  In this paper, we now
specialize to discuss the QCD domain walls presented in
\cite{Forbes:2000et} showing that they may be able to generate
sufficiently large magnetic fields to seed galactic dynamos and
possibly observe in the extra-galactic medium.

\subsection{Strong CP Violations in Domain Walls}
\label{sec:CP}
We present a brief summary here of the essence of strong CP violation
to explain how axion and QCD domain walls may satisfy the criteria
discussed above.  The most general form for the fundamental QCD
Lagrangian is known to contain the following term related to the
anomaly:
\begin{equation}
  \mathcal{L}_{\theta}=
  \frac{\theta g^2}{32\pi^2}\;G_{\mu\nu}^a\tilde{G}^{a\mu\nu}
\end{equation}
where $G_{\mu\nu}^a$ is the gluon field tensor and
$\tilde{G}^{a\mu\nu}=\frac{1}{2}\varepsilon^{\mu\nu\rho\sigma}
G_{\rho\sigma}^a$ is its dual.  This term is odd under the discrete
symmetry CP, thus, if $\theta$ is non-zero, then the strong
interaction should violate CP.  Experiments, however, have placed
tight limits $|\theta|<10^{-9}$.  The contributions to the final
$\theta$ arise from several sources, and it remarkable that these seem
to exactly cancel.  The origin of this cancellation is known as the
strong CP problem.

One solution is to promote $\theta$ from a parameter to a dynamical
field called the axion
\cite{Peccei:1977hh,Weinberg:1978ma}
.  The idea is
that, prior to the QCD phase transition, the axion field is massless
and $\theta$ can take on any value.  After the transition, the axion
acquires a mass and sits in a potential with a minimum energy where
$\theta=0$.  The axion field thus relaxes to the minimum restoring CP
conservation today.  To date, axions have not been detected, however,
there is an allowed region consistent with experimental, astrophysical
and cosmological constraints: Thus, the so-called invisible axion
\cite{Kim:1979if,Dine:1981rt}
, which
is very light: $m_a\sim 10^{-5}$--$ 10^{-3}$ eV, may still resolve the
strong CP problem.  In addition, axions of this mass are very strong
cold dark candidates (see for example the recent review
\cite{Turner:1999xj}).  As mentioned, axions provided a nice mechanism
for generating domain walls, but because the axion must be so light,
there is no way for such structures to efficiently interact with
nucleons.  For a good reviews of the strong CP problem and the role of
axions, see \cite{Cheng:1988gp,Peccei:1989,Kim:1999ia}.

In any case, we assume that some method exists to solve the strong CP
problem. What is important about the $\theta$ parameter is that, in
the low-energy limit, it only appears in the combinations
$(\theta+\phi)$ (Equation \ref{eq:V}) and $(\theta+\phi-a)$, where
$\phi$ is the dynamical field related to the $\eta'$ meson and $a$ is
the axion field.  Thus, even when $\theta=0$, CP will be violated in
strong interactions in a domain wall background where $\phi$ or
$\phi-a$ is non-zero over a macroscopically large region.  Hence, QCD
and axion domain walls induce strong CP violations over their central
regions.

One of the consequences of this strong CP violation is that nucleons
have an induced electric dipole moment as well as a magnetic dipole
moment\footnote{Normally the electric dipole moment is suppressed to
  the same order as $\theta$ in that CP is conserved.}
\cite{Crewther:1979pi}.  We summarize those results here.  

In the chiral limit $m_q \rightarrow 0$ for small $\theta$
\begin{equation}
  \label{eq:em9}
  d_N \simeq \frac{e g_{\pi NN} \bar{g}_{\pi NN} }{4\pi^2 m_N}
  \ln\left(\frac{m_N}{m_{\pi}}\right),
\end{equation}
where $g_{\pi NN}$ is the strong $\pi NN$ coupling constant and
$\bar{g}_{\pi NN}$ is the $CP$ odd $\pi NN$ coupling constant which
was estimated to be $\bar{g}_{\pi NN} \sim 0.04|\theta|$.  In these
formulae the $\theta$ parameter should be treated as the singlet
$\phi$ domain wall solution $\phi(z)$ with nontrivial $z$ dependence.
From these formulae one can compute the following relation
\begin{equation}
  \label{eq:em10}
  \frac{d_{\Psi}}{\mu_{\Psi}}\sim\frac{g_{\pi NN} m_q}{2\pi^2f_{\pi}}
  \ln\left(\frac{m_N}{m_{\pi}}\right)\theta(z)\simeq 0.1.
\end{equation}
Thus, for all nucleons, including the neutron, both the electric and
magnetic dipole moments are non-zero and of the same order in the
domain wall background when $\theta(z)\equiv\phi(z)\sim 1$.

\section{QCD Domain Walls}
\label{sec:QCDdw}
We saw in section~(\ref{sec:DWsources}) that several types of domain
walls might act as sources for seed fields. To be concrete, we shall
now restrict our attention to QCD domain walls to show how domain
walls might produce magnetic seed fields.  In this section, we shall
present a short review of the results presented in
\cite{Forbes:2000et}, simplifying the model for presentation.

To describe these walls, we consider the low-energy effective theory
of QCD including the pions and the $\eta'$ singlet field.  The $\eta'$
field is not as light but is the source of the physics behind the QCD
domain walls.  The pions and $\eta'$ enter the Lagrangian through the
matrix representation
\begin{equation}
  \label{eq:Udef}
  \mathbf{U}=\exp \left[ i \sqrt{2} \, \frac{\pi^{a} \lambda^{a} }{f_{\pi}}  + 
    i \frac{ 2}{ \sqrt{N_{f}} } \frac{ \eta'}{ f_{\eta'}}\right]
\end{equation}
where $\pi^a$ are the $N_f^2-1$ pseudo-Goldstone fields, $\lambda^a$
are the Gell-Mann matrices for $SU(N_f)$ and $\eta'$ is the singlet
field.  From now on, we limit ourselves to the simplest case of one
flavour $N_f=1$ which contains only the $\eta'$ field but captures all
the relevant physics.  The $N_f=2$ case is presented in
\cite{Forbes:2000et}.  Although the models are quantitatively
different, the phenomena described by both is the same.  In this
model, we see that (\ref{eq:Udef}) reduces to a single complex phase
\begin{align}
  \mathbf{U}&=e^{i\phi},&
  \phi=\frac{2\eta'}{f_{\eta'}}.
\end{align}

The effective Lagrangian density then reduces to
\begin{equation}
  \label{eq:Leff}
  \mathcal{L} = \frac{f_{\eta'}^2}{8}(\partial_\mu\phi)^2
  -V(\phi)
\end{equation}
with the effective potential
\begin{equation}
  \label{eq:V}
  V(\phi) = - \min_{l}\left\{
    M\cos\phi+E\cos\left(\frac{\theta+\phi+2\pi l}{N_c}\right)\right\}
\end{equation}
which was first introduced in \cite{Halperin:1998rc}.  The
minimization on the right comes from choosing the lowest energy branch
of the multi-valued potential.  Details about this potential are
discussed in the original paper \cite{Halperin:1998rc} but several
points will be made here.  All dimensionful parameters are expressed
in terms of the QCD chiral and gluon vacuum condensates: ${\bf M} =
{\rm diag} (m_{q}^{i} |\langle \bar{q}^{i}q^{i} \rangle| )$ is the
mass matrix and $ E = \langle b \alpha_s /(32 \pi) G^2 \rangle$ is the
vacuum energy.  These are well known numerically: $m_q\sim 5$ MeV (see
for example \cite{Groom:2000in}), $\langle\bar{q}q\rangle \sim
-(240\text{ MeV})^3$ \cite{Gasser:1982ap}, $\langle\alpha_{s}/\pi
G^2\rangle\sim 0.012$ GeV$^4$ \cite{Shifman:1979bx} and
$b=11N_c/3-2N_f/3$ is the first term in the beta function.

This potential correctly reproduces the Di~Vecchia-Veneziano-Witten
effective chiral Lagrangian in the large $N_c$ limit
\cite{Witten:1980sp,DiVecchia:1980ve}, it reproduces the anomalous
conformal and chiral Ward identities of QCD, and it reproduces the
known dependence in $\theta$ for small angles
\cite{Witten:1980sp,DiVecchia:1980ve}.  It also exhibits the correct
$2\pi$ periodicity in $\theta$.  This periodicity is the most
important property of the potential and is the reason that QCD domain
walls form: The qualitative results do not depend on the exact form
(cosine) of the potential.  Rather, the domain walls form naturally
because of the $2\pi$ periodicity ($\theta\rightarrow\theta+2\pi$)
which represents the discrete nature of the ground state symmetries.
It is exactly these symmetries which leads to the existence of axion
domain walls when $\theta$ is promoted to a dynamical axion field
\cite{Huang:1985tt,Chang:1998tb,Vilenkin:1982ks,Sikivie:1982qv,Lyth:1992tw}.

As described in section~(\ref{sec:CP}), we see that the singlet $U(1)$
field $\phi$ occurs in the same place as the CP violating $\theta$
parameter.  Thus, even though to a high degree of precision,
$\theta=0$, in the macroscopic regions where $\langle\eta'\rangle \neq
0$ is non-zero, there will be CP violating physics.

We see that the potential(\ref{eq:V}) has ground states characterized
by
\begin{equation}
  \label{eq:vac}
  \phi_0=2\pi n,
\end{equation}
where $n$ is an integer, with vacuum energy $V_{\mathrm{min}}=-M-E$.
Expanding about the minimum $\phi=\phi_0+\delta_{\phi}$ we find the
mass of the field\footnote{In the more general case of $N_f$ quarks
  with equal masses, the right-hand side of Equation (\ref{eq:mass})
  should be multiplied by a factor $N_f$.}
\begin{equation}
  \label{eq:mass}
  m_{\eta'}^2=\frac{4}{f_{\eta'}^2}\left(\frac{E}{N_c^2}+M\right)
\end{equation}
The most important point to realize is that all of the ground states
(\ref{eq:vac}) in fact represent the same physical state
$\mathbf{U}=\mathbf{1}$.  Thus, it is possible for the $\phi$ field to
make a transition $2\pi n\rightarrow2\pi m$ for different integers $n$
and $m$.  Within this model (\ref{eq:Leff}), where all heavy degrees
of freedom have been integrated out, these transitions are absolutely
stable and represent the domain walls.  When one includes the effects
of the heavier degrees of freedom, however, we find that the walls are
unstable on the quantum level.  This is described in detail in
\cite{Forbes:2000et} and briefly reviewed in the next section.

\subsection{Domain Wall Solutions}
To study the structure of the domain wall we look at a simplified
model where one half of the universe is in one ground state and the
other half is in another.  The fields will orient themselves in such a
way as to minimize the energy density in space, forming a domain wall
between the two regions.  In this model, the domain walls are planar
and we shall neglect the $x$ and $y$ dimensions: A complete
description of this wall is given by specifying the boundary
conditions and by specifying how the fields vary along $z$.

We present here the two basic domain wall solutions.  These are
characterized by interpolations from the state $\phi = 0$ to:
\begin{description}
  \item[Soliton:] $\phi=2\pi$,
  \item[Anti-soliton:] $\phi=-2\pi$.
\end{description}
It is possible to consider transitions between further states (i.e.
$0\rightarrow2\pi n$) but these can be thought of as multiple domain
walls.  They also have higher energies, are less stable, and are thus
less important for our discussion.  To gain an understanding of the
structure of the domain walls we look for the solution which minimizes
the energy density of the domain wall.  The energy density (wall
tension) per unit area is given by the following expression
\begin{equation}
  \label{eq:tension}
  \sigma = \int_{-\infty}^{\infty}\left(
    \frac{f_{\eta'}^2}{8}\dot{\phi}^2+
    V(\phi)-V_{\rm min}
  \right){\rm d}z
\end{equation}
where the first term is the kinetic contribution to the energy and the
last term is the potential.  Here, a dot signifies differentiation
with respect to $z$: $\dot{a}=\frac{{\rm d}a}{{\rm d}z}$.

To minimize the wall tension (\ref{eq:tension}), we can use the
standard variational principle to arrive at the following equations
of motion for the domain wall solutions:
\begin{equation}
  \label{eq:eqm}
  \frac{\ddot{\phi} f_{\eta'}^2}{4M} =
  \sin\phi + \frac{E}{M N_c}\sin\frac{\phi}{N_c}.
\end{equation}
Again, the last term of Equation (\ref{eq:eqm}) should be understood
as the lowest branch of a multi-valued function as described by
Equation (\ref{eq:Leff}).

\begin{figure}[ht]
  \begin{centering}
    \psfrag{eta}{$\phi$}
    \psfrag{cz}{$z\mu$}
    \includegraphics[width=0.4\textwidth]{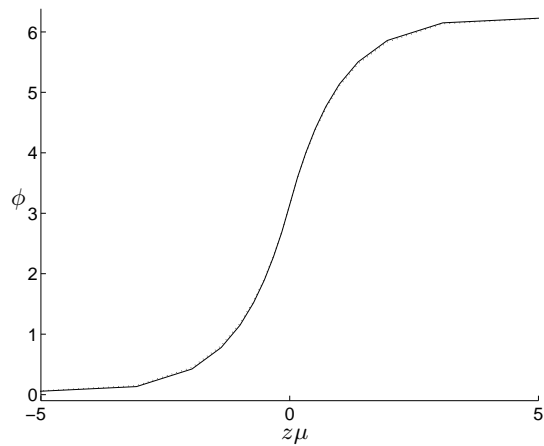}
    \caption{Basic form of the QCD domain wall (soliton).  The
      analytic approximation (\ref{eq:QCDetasol}) is plotted as a dotted
      line to show the good agreement.  We have taken $N_c=3$ here.
      Notice that the wall thickness is set by the parameter $\mu$.
      \label{fig:QCDwall}}
  \end{centering}
\end{figure}

The general analytical solution of Equations (\ref{eq:eqm}) is not
enlightening and we present the numerical solution in
Fig.~\ref{fig:QCDwall}.  In order to gain an intuitive understanding
of this wall, we examine the solution in the limit $M\ll
E/N_c^2$ (physically, when $N_f>1$, this is the limit $m_{\pi}\ll
m_{\eta'}$).  In this case, the last term of (\ref{eq:eqm}) dominates.
Thus, the structure of the $\phi$ field is governed by the
differential equation:
\begin{equation}
  \label{eq:QCDeta}
  \ddot{\phi}
  =\frac{4E}{N_c f_{\eta'}^2}\sin\frac{\phi}{N_c}.
\end{equation}
Now, there is the issue of the cusp singularity when $\phi=\pi$
because we change from one branch of the potential to another (see
Equation (\ref{eq:Leff}).) By definition, we keep the lowest energy
branch, such that the right-hand side of Equation (\ref{eq:QCDeta}) is
understood to be the function $\sin(\phi/N_c)$ for $ 0 \leq \phi \leq
\pi $ and $\sin([\phi -2\pi]/N_c)$ for $\pi \leq \phi \leq 2\pi$.
However, we notice that the equations of motion are symmetric with
respect to the center of the wall (which we take as $z=0$), hence
$\phi=\pi$ only at the center of the wall and not before, so we can
simply look at half of the domain, $z\in(-\infty,0]$, with boundary
conditions $\phi(-\infty)=0$ at $z=-\infty $ and $\phi(0)=\pi$ at
$z=0$.  The rest of the solution will be symmetric with $\phi=2\pi$
at $z=+\infty$.

Equation (\ref{eq:QCDeta}) with the boundary conditions above has the
solution
\begin{equation}
  \label{eq:QCDetasol}
  \phi(z) = 
  \begin{cases}
    4N_c \tan^{-1}\left[
      e^{\mu z}
      \tan\frac{\pi}{4 N_c} 
    \right],
    &  z\leq 0,  \\   
    2\pi - 4N_c\tan^{-1} \left[
      e^{-\mu z}
      \tan\frac{\pi}{4 N_c} 
    \right],
    & z \geq 0.
  \end{cases}
\end{equation}
which is a good approximation of the solution to Equation (\ref{eq:eqm})
when $M\ll E/N_c^2$.  Here, the scale of the wall is set by the
parameter $\mu$:
\begin{equation}
  \label{eq:QCDwidth}
  \mu \equiv \frac{2 \sqrt{E}}{ N_c f_{\eta'} }, 
  ~~~ \lim_{m_q\rightarrow 0}\mu = m_{\eta'},
\end{equation} 
which is the inverse width of the wall and which is equal to the field
mass $m_{\eta'}$ in the chiral limit $m_q^2\rightarrow 0$ (see
Equation (\ref{eq:mass})).  Thus, we see that, indeed, the QCD domain
walls have a QCD scale.

Solution (\ref{eq:QCDetasol}) describes the soliton.  The anti-soliton
can be found by taking $z\rightarrow-z$: thus, we have the transition
soliton$\rightarrow$anti-soliton under the discrete CP symmetry.  The
numerical solution for the $\phi$ field is shown in
Fig.~\ref{fig:QCDwall}.  It turns out that the approximation is
reasonable even in the physical case where $N_c^2M/E \sim 10^{-1}$.

The wall surface tension defined by Equation (\ref{eq:tension}) and
can be easily calculated analytically in the chiral limit when the
analytical solution is known and is given by Equation
(\ref{eq:QCDetasol}).  Simple calculations leads to the following
result:\footnote{In the general case of $N_f$ quarks of equal mass, the
  right-hand side of Equation (\ref{eq:tensionnum}) should be
  multiplied by the factor $1/\sqrt{N_f}$.  In this case, it reduces
  to the result cited in \cite{Forbes:2000et} when $N_f=2$.}
\begin{equation}
  \label{eq:tensionnum}
  \sigma=4 N_c  
  f_{\eta'}\sqrt{\left\langle \frac{ b \alpha_s }{ 32 \pi} G^2 \right\rangle }
  \, \left( 1 - \cos \frac{\pi}{ 2 N_c} \right) + {\rm O}(m_q
  f_{\eta'}^2 ).
\end{equation} 
In the case when $m_q\neq 0$, $\sigma$ is numerically close to the
estimate (\ref{eq:tensionnum}).

\subsection{Domain Wall Decay}
Finally, we note that these domain walls are not stable: as mentioned
earlier, the vacuum states (\ref{eq:vac}) represent the same physical
state.  When one includes the heavier gluonic degrees of freedom, it
becomes possible for the fields to ``unwind'' through this extra
degree of freedom.  Classically this is not allowed because the heavy
degrees of freedom are constrained by a large potential barrier, but
it is still possible for the field to tunnel through this barrier
forming a hole in the domain wall.  Once a large enough hole is
formed, it will expand and consume the domain wall.  This process is
called ``nucleation'' and is similar to the mechanism consuming $N=1$
axion domain walls
\cite{Vilenkin:1994paper,Chang:1998tb,Vilenkin:1982ks,Kibble:1982dd}.

In \cite{Forbes:2000et}, we estimate the lifetime of these domain
walls borrowing the same methods used to estimate the lifetime of
axion domain walls in the $N=1$ axion models
\cite{Kibble:1982dd,Vilenkin:1982ks,Vilenkin:1994paper,Chang:1998tb}.  We
should point out one major difference, however, between the $N=1$
axion model and our model.  In the axion model, prior to domain wall
formation there is a phase where stable axion strings can form.  When
the domain walls form, these strings bound the domain walls and thus
the walls start to decay from the outset greatly reducing their
lifetime (see \cite{Vilenkin:1994paper} for a nice discussion).  In the
case of QCD domain walls, strings are not stable objects and do not
form before the walls.  Thus, the only way for the walls to decay is
through the nucleation process we are about to describe.  This greatly
suppresses the decay rate and is the source of the long lifetime for
the walls.

The tunneling probability can be estimated by computing the action
$S_0$ of an instanton solution of the Euclidean (imaginary time,
$t=i\tau$) field equations, approaching the unperturbed wall solution
at $\tau\rightarrow \pm\infty$.  In this case the probability $P$ of
creating a hole is proportional to the factor
\begin{equation}
  \label{d3}
  P \sim e^{-S_0}
\end{equation}
where $S_0$ is the classical instanton action.

If the radius $R_c$ of the nucleating hole is much greater than the
wall thickness, we can use the thin-string and thin-wall
approximation.  In this case, the action for the string and for the
wall are proportional to the corresponding world-sheet areas
\cite{Kibble:1982dd},
\begin{equation}
  \label{eq:action}
  S_0=4\pi R^2\alpha-\frac{4\pi}{3}R^3\sigma .
\end{equation}
Here $\sigma$ is the wall tension (\ref{eq:tensionnum}), and $\alpha \sim
\sqrt{2E}$ is the string tension which we estimate based on
dimensional arguments.  The string tension $\alpha$ tries to close the
hole while the wall tension $\sigma$ tries to widen the hole.
Minimizing (\ref{eq:action}) with respect to $R$ we find that
\begin{align}
  \label{eq:d5}
  R_c &= \frac{2\alpha}{\sigma}, & S_0 &= \frac{16\pi\alpha^3}{3\sigma^2}.
\end{align}
If a hole forms with radius $R>R_c$ then the hole will expands with
time as $x^2+y^2=R^2+t^2$, rapidly approaching the speed of light and
consuming the domain wall.

Inserting numerical values for the phenomenological relevant case
$N_f=2$ we find that \cite{Forbes:2000et}
\begin{align}
  \label{d8}
  \alpha &\sim (0.28\text{ GeV})^2, &
  \sigma &= (200\text{ MeV})^3, &
  S_0 &\sim 120.
\end{align}
What is important is that $S_0$ is numerically large, and hence the
lifetime is much larger than the QCD scale because of the huge
tunneling suppression $e^{-S_0}\sim 10^{-52}$.  A more complete analysis
is presented in \cite{Forbes:2000et} where we estimate the lifetime of
the walls to be of the order
\begin{equation}
  \label{eq:DWtau}
  \tau \sim 10^{-5} \text{ s}
\end{equation}
even though the walls are governed by the microscopic QCD scale.  This
result should be interpreted with some caution: in the low 
energy regime, we do not have very good control over the quantitative
physics.\footnote{It is possible to regain control of the calculation
  in the high density limit.  See the discussion in
  section~(\ref{sec:Summary}).}
Arguments presented in \cite{Forbes:2000et}
 show, however, that it is at least possible for domain
walls of purely QCD origin to live for macroscopically large
lifetimes.

To summarize, we have a QCD domain wall with all of the properties
required to generate magnetic fields:
\begin{enumerate}
\item The walls form shortly after the QCD phase transition and attain 
  Hubble-scale correlations through the Kibble mechanism.
\item The QCD domain walls have a structure on the scale of
  $m_{\eta'}^{-1}\sim\Lambda_{\mathrm{QCD}}^{-1}$ and thus they can
  efficiently interact with nucleons and other QCD matter.
\item The transition in the singlet $\eta'$ field produces an
  environment near the wall where the effective $\theta$ parameter is
  non-zero.  Thus, across the wall, there is maximal CP violation.  In
  such an environment, it is known that the electric and magnetic
  dipole moments of the nucleons are of the same order
  \cite{Crewther:1979pi}.  
\item The strong CP violation also provides a mechanism for generating
  helicity on a Hubble scale\footnote{To be precise, the domain walls
    separate the helicity into Hubble size regions to that globally
    the total helicity is zero, but within Hubble scale regions, the
    helicity is maximal and correlated with the same sign.} by
  aligning both the electric and magnetic dipole moments along the
  domain wall.
\item The decay mechanism described in \cite{Forbes:2000et} renders
  the QCD domain walls unstable such that the walls themselves do not
  pose a cosmological problem.  However, the suppression in the decay
  mechanism due to quantum tunneling might extend the lifetime of the
  walls to a macroscopic scale (\ref{eq:DWtau}) which is long enough
  to generate the required electromagnetic turbulence as we shall
  show.
\end{enumerate}

\section{Alignment of Spins in the Domain Wall}
\label{sec:align}
Now we shall show that the domain walls indeed do acquire a
magnetization and present a simplified method for estimating the
magnitudes of bulk properties on the domain wall.  This method makes
the approximation that the domain wall is flat and that translational
and rotational symmetries are preserved in the plane of the wall (which
we take to be the $x$--$y$ plane.)  These approximations are valid in
the case of domain walls whose curvature is large in comparison to the
length scale of the pertinent physics.

Once this approximation is made, we can reformulate the problem in
$1+1$ dimensions ($z$ and $t$) and calculate the density of the
desired bulk properties along the domain wall.  To regain the full
four-dimensional bulk properties, we must estimate the density of the
particles in the $x$--$y$ plane to obtain the appropriate density and
degeneracy factors for the bulk density.  Thus, the final results are
not independent of physics in the $x$--$y$ plane, but rather, these
effects are accounted for only through the degeneracy factors.

We proceed to demonstrate this technique by calculating the alignment
of fermionic spins along the wall.  We take the standard form for the  
interaction between the pseudo-scalar $\eta'$ field and the nucleons
which respects all relevant symmetries:
\begin{equation}
  \label{eq:L4d}
  \mathcal{L}_4=\bar{\Psi}\left(i{\not{\!\partial}}-
    m_Ne^{i\phi\gamma_5}\right)\Psi.
\end{equation}
Here $\phi=\phi(z)$ characterizes our domain wall solution as
expressed in Equation (\ref{eq:QCDetasol}) and $m_N$ is the nucleon
mass.  For our approximations, we assume that fluctuations in the
nucleon field $\Psi$ do not affect the domain walls and, thus, treat
the domain walls as a background field\footnote{A full account would
  take into account the effects of this back-reaction.  We expect that
  such back-reactions would affect the potential (\ref{eq:V}) by
  altering the form of the last term and possibly adding higher order
  corrections.  This may affect the magnitudes of some of the
  estimates, but would certainly not alter the topology of the fields
  and thus the domain walls would still form with a similar structure.
  Quantitatively this would alter the results, but not the order of
  magnitude.}.  The strategy is to break (\ref{eq:L4d}) into two $1+1$
dimensional components by setting $\partial_x=\partial_y=0$ (this is
the approximation that the physics in the $z$ direction decouples from
the physics in the $x$--$y$ plane) and then by manipulating the system
of equations that result.

First, we introduce the following chiral components of the Dirac
spinors \footnote{We are using the standard representation here:
\begin{align*}
  \gamma_0&=\begin{pmatrix}
    {I} & {0}\\
    {0} & -{I}
  \end{pmatrix},&
  {\gamma}_j&=\begin{pmatrix}
    {0} & {\sigma}_j\\
    -{\sigma}_j & {0}
  \end{pmatrix},&
  {\gamma}_5&=\begin{pmatrix}
    {0} & {I}\\
    {I} & {0}
  \end{pmatrix},
\end{align*}
\begin{align*}
  \sigma_1&=\begin{pmatrix}
    0 & 1\\
    1 & 0
  \end{pmatrix},&
  \sigma_2&=\begin{pmatrix}
    0 & -i\\
    i & 0
  \end{pmatrix},&
  \sigma_3&=\begin{pmatrix}
    1 & 0\\
    0 & -1
  \end{pmatrix}.
\end{align*}
}
:
\begin{align}
  \Psi_+&=\frac{1}{\sqrt{S}}\begin{pmatrix}
    \chi_1\\
    \chi_2
  \end{pmatrix}, &
  \Psi_-&=\frac{1}{\sqrt{S}}\begin{pmatrix}
    \xi_1\\
    \xi_2
  \end{pmatrix},
\end{align}
\begin{equation}
  \Psi=\frac{1}{\sqrt{2S}}\begin{pmatrix}
    \chi_1+\xi_1\\
    \chi_2+\xi_2\\
    \chi_1-\xi_1\\
    \chi_2-\xi_2
  \end{pmatrix}=
  \frac{1}{\sqrt{2}}\begin{pmatrix}
    \Psi_++\Psi_-\\
    \Psi_+-\Psi_-
  \end{pmatrix},
\end{equation}
where $S$ is the area of the wall. This normalization factor cancels
the degeneracy factor proportional to $S$ added in the next subsection.
Now we re-express (\ref{eq:L4d}) by noting that ${\gamma}_5^2={I}$:
\begin{equation*}
  \mathcal{L}_4=\bar{\Psi}\left[i\begin{pmatrix}
      \partial_0 & -{\sigma}_j\partial_j\\
      {\sigma}_j\partial_j & -\partial_0
    \end{pmatrix}
    -m_N\begin{pmatrix}
      \cos(\phi) & i\sin(\phi)\\
      i\sin(\phi) & \cos(\phi)
    \end{pmatrix}    
  \right]\Psi.
\end{equation*}
The associated Dirac equation is
\begin{equation*}
  \left[i\begin{pmatrix}
    \partial_0 & -{\sigma}_j\partial_j\\
    {\sigma}_j\partial_j & -\partial_0
  \end{pmatrix}
    -m_N\begin{pmatrix}
      \cos(\phi) & i\sin(\phi)\\
      i\sin(\phi) & \cos(\phi)
    \end{pmatrix}
  \right]\Psi=0.
\end{equation*}
This is equivalent to the coupled system:
\begin{subequations}
\begin{align}
  2i(\partial_0+\sigma_i\partial_i)\Psi_-&=2m_Ne^{i\phi}\Psi_+,\\
  2i(\partial_0-\sigma_i\partial_i)\Psi_+&=2m_Ne^{-i\phi}\Psi_-.
\end{align}
\end{subequations}
Now, we decouple the $z$ coordinates from $x$ and $y$ by setting
$\partial_x=\partial_y=0$:
\begin{subequations}
  \label{eq:decouple}
  \begin{align}
    i\begin{pmatrix}
      \partial_t+\partial_z&0\\
      0&\partial_t-\partial_z
    \end{pmatrix}
    \Psi_-&=m_Ne^{i\phi}\Psi_+,\\
    i\begin{pmatrix}
      \partial_t-\partial_z&0\\
      0&\partial_t+\partial_z
    \end{pmatrix}\Psi_+&=m_Ne^{-i\phi}\Psi_-.
  \end{align}
\end{subequations}
Both these equations are diagonal.  Thus, we see that the top
components and bottom components of $\Psi_\pm$ mix independently.
\begin{subequations}
  \begin{align}
    \begin{pmatrix}
      -m_Ne^{i\phi}&i(\partial_t+\partial_z)\\
      i(\partial_t-\partial_z)&-m_Ne^{-i\phi}
    \end{pmatrix}
    \begin{pmatrix}
      \chi_1\\
      \xi_1
    \end{pmatrix}=0,\\
    \begin{pmatrix}
      -m_Ne^{i\phi}&i(\partial_t-\partial_z)\\
      i(\partial_t+\partial_z)&-m_Ne^{-i\phi}
    \end{pmatrix}
    \begin{pmatrix}
      \chi_2\\
      \xi_2
    \end{pmatrix}=0.
  \end{align}
\end{subequations}
Remember that we are looking for a two-dimensional Dirac equation,
thus we want the kinetic terms to look the same.  For this reason we
should flip the rows and columns of the second equation.  Doing this
and defining the two two-dimensional spinors
\begin{align}
  \Psi_{(1)}&=\begin{pmatrix}
    \chi_1\\
    \xi_1
  \end{pmatrix},&
  \Psi_{(2)}&=\begin{pmatrix}
    \xi_2\\
    \chi_2
  \end{pmatrix},
\end{align}
the equations have the following structure:
\begin{subequations}
  \label{eq:D2d}
  \begin{align}
    (i{\hat{\gamma}}^\mu\partial_\mu
    -m_Ne^{+i\phi{\hat{\gamma}}_5})\Psi_{(1)}&=0
    \\
    (i{\hat{\gamma}}^\mu\partial_\mu
    -m_Ne^{-i\phi{\hat{\gamma}}_5})\Psi_{(2)}&=0
  \end{align}
\end{subequations}
where the index $\mu\in\{t,z\}$, the Lorentz signature is $(1,-1)$ and
we define the following two-dimensional version of the gamma matrices:
\begin{align*}
  {\hat{\gamma}}_t&={\sigma}_1&,
  {\hat{\gamma}}_z&= -i{\sigma}_2, &
  {\hat{\gamma}}_5&= {\sigma}_3.
\end{align*}
These satisfy the proper two-dimensional relationships
$\hat{{\gamma}}_5 = \hat{{\gamma}}_t\hat{{\gamma}}_z$ and
${\hat{\gamma}}_\mu{\hat{\gamma}}_\nu =
g_{\mu\nu}+\epsilon_{\mu\nu}{\hat{\gamma}}_5$.  We can reproduce
equation (\ref{eq:D2d}) from the following effective two-dimensional
Lagrangian density,
\begin{align}
  \mathcal{L}_2=&\bar\Psi_{(1)}\left(i{\hat{\gamma}}^\mu\partial_\mu
    -m_Ne^{+i\phi{{\hat{\gamma}}_5}}\right)\Psi_{(1)}+\nonumber\\
  +&\bar\Psi_{(2)}\left(i{\hat{\gamma}}^\mu\partial_\mu
    -m_Ne^{-i\phi{{\hat{\gamma}}_5}}\right)\Psi_{(2)},
  \label{eq:L2d}
\end{align}
where two different species of fermion with opposite chiral charge
interact with the domain wall background $\phi(z)$.  Note that, due to
the normalization factor $1/\sqrt{S}$ we introduced above, the
two-dimensional fields $\Psi_{(i)}$ have the correct canonical
dimension $1/2$.

We have thus successfully reduced our problem to a two-dimensional
fermionic system.  It is known that for several systems in $1+1$
dimensions, the fermionic representation is is equivalent to a $1+1$
dimensional bosonic system through the following
equivalences\footnote{The constant $\mu$ in the last two equations is
  a scale parameter of order $m_N$.  The exact coefficient of this
  term depends on the model an is only known for exactly solvable
  systems but in all cases, is of order unity. This technique is
  well-known to the condensed matter and particles physics
  communities. See for example \cite{Goldstone:1981kk,Jackiw:1976fn}.}
\cite{Mandelstam:1975hb,Coleman:1975bu}:
\begin{subequations}
  \begin{align}
    \bar\Psi_{(j)}i{\hat{\gamma}}^\mu\partial_\mu\Psi_{(j)}
    &\rightarrow
    \frac{1}{2}(\partial_\mu\theta_j)^2,\\
    \bar\Psi_{(j)}{\hat{\gamma}}_\mu\Psi_{(j)}
    &\rightarrow
    \frac{1}{\sqrt{\pi}}\epsilon_{\mu\nu}\partial^\nu\theta_j,\\
    \bar\Psi_{(j)}\Psi_{(j)}
    &\rightarrow
    -\mu\cos(2\sqrt{\pi}\theta_j),\\
    \bar\Psi_{(j)}i{\hat{\gamma}}_5\Psi_{(j)}
    &\rightarrow
    -\mu\sin(2\sqrt{\pi}\theta_j).
  \end{align}
\end{subequations}
After making these replacements, we are left with the following
two-dimensional bosonic effective Lagrangian density describing the
two fields $\theta_1$ and $\theta_2$ in the domain wall background
$\phi(z)$
\begin{equation}
  \mathcal{L}=\frac{1}{2}(\partial_\mu\theta_1)^2
  +\frac{1}{2}(\partial_\mu\theta_1)^2
  -U(\theta_1,\theta_2)
\end{equation}
where the effective potential is
\begin{equation}
  \label{eq:Ubose}
  U(\theta_1,\theta_2)=-m_N\mu\left[\cos(2\sqrt{\pi}\theta_1-\phi)
    +\cos(2\sqrt{\pi}\theta_2+\phi)\right].
\end{equation}
The next approximation that we make is to neglect the dynamics of the
$\theta_k$ fields: we assume that they relax slowly in the domain wall
background such that their dynamics do not contribute appreciably to the
final state\footnote{This is the same adiabatic approximation used by
  Goldstone and Wilczek \cite{Goldstone:1981kk}.} which minimizes the
potential (\ref{eq:Ubose}).  The classical minimizing solution is thus
\begin{align}
  \label{eq:2dSol}
  \langle\theta_1\rangle &= \frac{\phi}{2\sqrt{\pi}}, &
  \langle\theta_2\rangle &= \frac{-\phi}{2\sqrt{\pi}}.
\end{align}
We are now ready to show that the domain walls align the spins of the
fermions.  The relevant spin operator is
 \footnote{Here we use the
   convention that ${\sigma}_{ij}=\frac{1}{2}[{\gamma}_i,{\gamma}_j]$,
   thus the spin operator
   ${\Sigma}_k\equiv\frac{i}{2}\epsilon_{ijk}{\sigma}_{ij}$.  In matrix
   form with the standard representation, this becomes:
   \begin{equation*}
     {\Sigma}_k=\epsilon_{ijk}\frac{i}{4}\begin{pmatrix}
       -[{\sigma}_i,{\sigma}_j] & 0\\
       0 & -[{\sigma}_i,{\sigma}_j]
     \end{pmatrix}
     =\begin{pmatrix}
       {\sigma}_k & 0\\
       0 & {\sigma}_k
     \end{pmatrix}.
   \end{equation*}
   In terms of the gamma matrices, this is
   ${\Sigma}_k={\gamma}_0{\gamma}_k{\gamma}_5$.}
\begin{equation}
  \Psi^{\dagger}\vec{{\Sigma}}\Psi
  = \bar{\Psi} \vec{{\gamma}} {\gamma}_5 \Psi 
  = \Psi_{+}^{\dagger} \vec{{\sigma}} \Psi_{+}
  + \Psi_{-}^{\dagger} \vec{{\sigma}}  \Psi_{-}
\end{equation}
Let us consider the $z$ component of the spin. We then have
\begin{align}
  \Psi^{\dagger}{\Sigma}_z\Psi
  &= \frac{1}{S}(\chi_1^\dagger\chi_1-\chi_2^\dagger\chi_2+
  \xi_1^\dagger\xi_1-\xi_2^\dagger\xi_2 )\nonumber\\
  &= \frac{1}{S}(\bar{\Psi}_{(1)}{\hat{\gamma}}_t\Psi_{(1)}-
  \bar{\Psi}_{(2)}{\hat{\gamma}}_t\Psi_{(2)})
  \label{eq:2dSpin}
\end{align}
and so we see that 
the four-dimensional spin operator
$\Sigma_z =  \gamma_0  \gamma_z  \gamma_5$ is expressed in terms
of a pair of two-dimensional fermion charge
operators.  We can
calculate the expectation value of the spin operator in the domain
wall background using this two-dimensional correspondence
(\ref{eq:2dSpin}) and the bosonic representation of the fermions
\begin{align}
  \Psi^{\dagger}{\Sigma}_z\Psi 
  &= \tfrac{1}{S}(\bar{\Psi}_{(1)}{\hat{\gamma}}_t\Psi_{(1)}
  - \bar{\Psi}_{(2)}{\hat{\gamma}}_t\Psi_{(2)}) \nonumber\\
  &= \tfrac{1}{S\sqrt{\pi}}\partial_z(\theta_2-\theta_1).
\end{align}
Finally, we use our minimizing bosonic solution (\ref{eq:2dSol}) to
obtain the following four-dimensional average spin aligned along the
domain wall:
\begin{equation}
  \label{eq:Spin}
  \langle{\Psi^{\dagger}{\Sigma}_z\Psi}\rangle
  =-\frac{1}{S\pi}\frac{\partial\phi(z)}{\partial z}.
\end{equation}
We will also need the following matrix elements later on:
\begin{align}
  \langle{\bar{\Psi}i{\sigma}_{xy}\Psi}\rangle
  &=\langle{\bar{\Psi}{\sigma}_{tz}{\gamma}_5\Psi}\rangle
  =\left\langle{\Psi^{\dagger}\begin{pmatrix}
      {\sigma}_3 & 0\\
      0 & -{\sigma}_3
    \end{pmatrix}
    \Psi}\right\rangle \nonumber\\
  &=\tfrac{1}{S}\langle{\chi_1^\dagger\xi_1-\chi_2^\dagger\xi_2+
    \xi_1^\dagger\chi_1-\xi_2^\dagger\chi_2}\rangle\nonumber\\
  &=\tfrac{1}{S}\langle{\bar{\Psi}_{(1)}\Psi_{(1)}}\rangle-
  \tfrac{1}{S}\langle{\bar{\Psi}_{(2)}\Psi_{(2)}}\rangle\nonumber\\
  &=0
  \label{eq:el1}
\end{align}
and
\begin{align}
  -\langle{\bar{\Psi}i{\sigma}_{tz}\Psi}\rangle
  &=\langle{\bar{\Psi}{\sigma}_{xy}{\gamma}_5\Psi}\rangle
  =\left\langle{\Psi^{\dagger}\begin{pmatrix}
      0 & -i{\sigma}_3 \\
      i{\sigma}_3 & 0
    \end{pmatrix}
    \Psi}\right\rangle\nonumber\\
  &=\tfrac{i}{S}\langle{-\xi_1^\dagger\chi_1+\xi_2^\dagger\chi_2+
    \chi_1^\dagger\xi_1-\chi_2^\dagger\xi_2}\rangle\nonumber\\
  &=-\tfrac{1}{S}\langle{\bar{\Psi}_{(1)}i\hat{{\gamma}}_5\Psi_{(1)}}\rangle+
 \tfrac{1}{S}\langle{\bar{\Psi}_{(2)}i\hat{{\gamma}}_5\Psi_{(2)}}\rangle
\nonumber\\
  &=\tfrac{2\mu}{S}\sin(\phi)
  \label{eq:el2}
\end{align}
Remember that we have restricted ourselves to a $1+1$ dimensional
theory.  We must now estimate the density and degeneracy of the
nucleons along the wall so we can obtain a true $1+3$ dimensional
estimate of the spin density.

\subsection{Fermion Degeneracy in the Domain Wall}
\label{sec:degeneracy}
We have assumed that locally the domain walls have only a spatial $z$
dependence.  There is still a two-dimensional translational and
rotational symmetry in the $x$--$y$ plane.  These translational
degrees of freedom imply that momentum in the plane is conserved and
hence we can treat the neglected degrees of freedom for the fermions
as free.  The degeneracy in a region of area $S$ will simply be a sum
over these degrees of freedom with a discrete factor $g=4=2\times2$
for spin and isospin degeneracy
\begin{equation}
  N=g\int_{\substack{\vec{x}\in S\\
      \lVert{\vec{p}}\rVert<p_F}}
  \mathrm{d}{x}\mathrm{d}{y}
  \frac{\mathrm{d}{p_x}}{2\pi}\frac{\mathrm{d}{p_y}}{2\pi}
  =g\frac{S\pi p_F^2}{4\pi^2}\sim g\frac{S\Lambda_{\mathrm{QCD}}^2}{4\pi}.
  \label{eq:Degen}
\end{equation}
Here we estimate the Fermi momentum $p_F\approx
\Lambda_{\mathrm{QCD}}$ by the thermal scale of the
fermions\footnote{We are treating the fermions along the wall as a
  massless, two-dimensional Fermi-gas.} and assume that the Fermi sea
is filled\footnote{Voloshin obtained a similar estimate
  \cite{Voloshin:2000jp}.  Furthermore, he estimates that there are
  sufficiently many fermions in the Hubble volume to diffuse into the
  domain wall potential justifying our assumption about the filled
  Fermi sea.}.

This completes our estimate of the induced spin along the domain wall
in a small region $S$.  Combining our estimate of the spin
(\ref{eq:Spin}) from the bosonization scheme with the fermion
degeneracy (\ref{eq:Degen}) we obtain the spin density along the wall:
\begin{equation}
  \langle{\Psi^{\dagger}{\Sigma}_z\Psi}\rangle_{4D}
  =\frac{\Lambda_{\mathrm{QCD}}^2}{\pi^2}\frac{\partial\phi(z)}{\partial z}.
\end{equation}
As a check, note that, the dimension here is $3$ and, the result does
not depend on the normalization factor $S$.
\section{Generation of Electromagnetic Field}
\label{sec:gen}
Once the spins are aligned, the nucleon electric and dipole moments
interact with the electromagnetic fields $F_{\mu\nu}$ through the
interaction
\begin{equation}
  \label{eq:Lem}
  \frac{1}{2}\left(
    d_{\Psi}\bar{\Psi}{\sigma}_{\mu\nu}{\gamma}_5\Psi 
    +\mu_{\Psi}\bar{\Psi}i{\sigma}_{\mu\nu}
    \Psi\right) F^{\mu\nu} + \bar{\Psi}(iD_\mu)(iD^\mu)\Psi .
\end{equation}
Here the nucleons have both electric and magnetic dipole moments
$d_{\Psi} \sim \mu_{\Psi}$ respectively (\ref{eq:em10}).

Now, we make the approximation again that the nucleons align
independently of the electromagnetic field, and we treat the nucleon
field $\Psi$ as a background.  The situation is a field of dipoles
aligned along the domain wall.  The net fields generated by surface of
area $\xi^2$ willed with a constant density of aligned dipoles is
proportional to $\xi^{-1}$ since the dipoles tend to cancel.  For a
perfectly flat domain wall of infinite extent, $\xi\rightarrow\infty$
and thus no net field would remain as pointed out in
\cite{Voloshin:2000jp}.  The QCD domain walls, however, are far from
flat: the walls have many wiggles and high frequency dynamics
excitations.  Thus, the fields generated by the dipoles will not
cancel on the domain wall, but will be suppressed by a
factor\footnote{The density of the dipoles is governed by the QCD
  scale $\Lambda_{\mathrm{QCD}}^{-1}$.} of $(\xi\Lambda_{\mathrm{QCD}})^{-1}$
where $\xi$ is an effective correlation length that depends on the
dynamics of the domain walls.  As an upper bound, the extent of the
domain walls is limited by the Hubble scale.  Typically, domain walls
remain space filling, thus we expect
$\Lambda_{\mathrm{QCD}}^{-1}\leq\xi\ll$~Hubble scale.  Unfortunately,
we presently cannot make a tighter bound on $\xi$, however, we shall
see that, even in the worst case, this mechanism can at least generate
feasible seed fields for galactic dynamos to amplify.

The result of sections~(\ref{sec:align}) and (\ref{sec:degeneracy}) is a
method for estimating the strengths of various sources in the domain
walls.  We now need to couple these to the generation of
electromagnetic turbulence.  To do this properly requires the solution
to Maxwell's equations as coupled to the sources in (\ref{eq:Lem}).
This is difficult, though no doubt important for accurate numerical
estimates, and so for an order of magnitude estimate we consider a
dimensional estimate considering the sources as a set of dipoles
sitting in the domain walls.  The spacing between the dipoles is set
by the QCD scale $\Lambda_{\mathrm{QCD}}^{-1}$, and the strengths 
of the   field can
be estimated from (\ref{eq:Lem}) using dimensional arguments:
\begin{equation}
\label{eq:F}
  \langle{F_{\mu\nu}}\rangle \sim
    \frac{1}{\xi\Lambda_{\mathrm{QCD}}}\left(
      d_{\Psi}\langle{\bar{\Psi}{\sigma}_{\mu\nu}{\gamma}_5\Psi}\rangle 
      +\mu_{\Psi}\langle{\bar{\Psi}i{\sigma}_{\mu\nu}\Psi}\rangle\right).
\end{equation}
This includes the dipole suppression discussed above.  
From (\ref{eq:el1}),   (\ref{eq:el2}), (\ref{eq:Degen}) and (\ref{eq:F}) 
we arrive at the following
estimates for the average  
electric and magnetic fields (correlated on the Hubble scale) which
includes the degeneracy factors (remember that
$\mu\approx m_N$):
\begin{equation}
  \label{eq:EB}
  |\langle E_z\rangle|\sim|\langle B_z\rangle|\sim
  \frac{0.5}{\pi}\frac{e\Lambda_{\mathrm{QCD}}^2}
  {\xi\Lambda_{\mathrm{QCD}}}\sim\frac{10^{17}\text{ G}} 
  {\xi\Lambda_{\mathrm{QCD}}}
\end{equation}

This method of estimating the electric and magnetic fields produced is
extremely crude: we have not solved Maxwell's equations, we have not
taken back reactions into account and we have not fully accounted for
the motion and geometry of the domain walls.  Never the less, we
expect that the estimates (\ref{eq:EB}) to be valid as an order of
magnitude estimate for the field strengths.  The approximations we
have made and effects that we have neglected will be discussed in
section~(\ref{sec:Summary}).  Thus, we have the approximations
for the fields (\ref{eq:EB}) which, 
along with (\ref{eq:scale}), justifies the estimate (\ref{eq:*}).

\subsection{Helicity}
\label{sec:helicity}
Finally, we note that the turbulence discussed in
section~(\ref{sec:evolution}) should be highly helical.  This helicity
arises from the fact that both electric and magnetic fields are
correlated together along the entire domain wall, $\langle\vec{\bf
  E}\rangle\sim\langle \vec{\bf A}\rangle/\tau$ where $\langle
\vec{\bf A}\rangle$ is the vector potential and $\tau$ is a relevant
timescale for the electrical field to be screened (we expect
$\tau\sim\Lambda_{\rm QCD}^{-1}$ as we discuss below).  The magnetic
helicity density is thus:
\begin{equation}
  \label{helicity}
  h\sim{\vec{\bf A}\cdot\vec{\bf B}} \sim\tau\langle E_z \rangle\langle
  B_z \rangle \sim \tau\frac{e^2}{\pi^2}\frac{\Lambda_{QCD}^2}{{\xi}^2}.
\end{equation}
It can be seen from (\ref{eq:F}) that both the electric field and
magnetic fields have the same structure in the domain wall.  This
implies that domain walls (solitons) have the same sign of helicity
everywhere.  Under CP the wall becomes an ``anti-wall'' (anti-soliton)
and the orientation of the magnetic field $B$ changes direction: thus
anti-walls have opposite helicity.

Note carefully what happens here: The total helicity was zero in the
quark-gluon-plasma phase and remains zero in the whole universe, but
the helicity is separated so that in one Hubble volume where one
domain wall dominates, the helicity has the same sign.  The reason for
this is that, as the domain walls coalesce, initial perturbations
cause either a soliton or an anti-soliton to dominate and fill the
Hubble volume.  In the neighboring space, there will be other solitons
and anti-solitons so that there is an equal number of both, but they
are separated and this spatial separation prevents them from
annihilating.  This is similar to how a particle and anti-particle may
be created and then separated so they do not annihilate.  In any case,
the helicity is a pseudo-scalar and thus maintains a constant sign
everywhere along the domain wall: thus, the entire Hubble volume is
filled with helicity of the same sign.  This is the origin of the
Hubble scale correlations in the helicity and in $B^2$.  The
correlation parameter $\xi$ which affects the magnitude of the fields
plays no role in disturbing this correlation.

As we mentioned, eventually, the electric field will be screened.  The
timescale for this is set by the plasma frequency for the electrons
(protons will screen much more slowly) $\omega_p$ which turns out to
be numerically close to $\Lambda_{\rm QCD}$ near the QCD phase
transition.  The nucleons, however, also align on a similar timescale
$\Lambda_{QCD}^{-1}$, and the helicity is generated on this scale too,
so the electric screening will not qualitatively affect the
mechanism.\footnote{If the screening were more efficient, then one might
  worry that the electric field would not last long enough to generate
  the helicity.}  Finally, we note that the turbulence requires a
seed which remains in a local region for a timescale set by the
conductivity \cite{Dimopoulos:1997nq} $\sigma\sim
cT/e^2\sim\Lambda_{\rm QCD}$ where for $T=100$ MeV, $c\approx 0.07$
and is smaller for higher $T$.  Thus, even if the domain walls move at
close to the speed of light (due to vibrations), there is still enough
time to generate turbulence throughout the Hubble volume.

\section{Conclusion}
\subsection{Summary}
\label{sec:Summary}
Here is a brief summary of what we have done and the approximations
that were made:

1) In section~(\ref{sec:DWsources}) we outlined the generic features
that domain wall models should posses if they are to successfully
generate primordial seeds by the method described in this paper.  We
proceeded with the QCD domain wall model \cite{Forbes:2000et} to
estimate the magnitude of the effects\footnote{
Further support for the existence of QCD domain walls comes from
calculations presented in \cite{Son:2000fh} where it is shown that
these walls almost certainly exist in the high density regime of QCD.
Thus, domain walls seem to be important features at high density.
Furthermore, if one accepts a conjecture on quark-hadron continuity at
low temperature with respect to variations in the chemical potential
$\mu$ \cite{Schafer:1998ef}, then one can make the following argument:
If for large $\mu$ domain walls exist, but for low $\mu$ they do not,
then there should be some sort of phase transition as one lowers
$\mu$.  Thus, the continuity conjecture supports the existence of
quasi-stable QCD domain walls at lower densities, at least down to the
densities of hyper-nuclear matter.  Coupled with the fact that gluon
and quark condensates do not vary much as one moves from the domain of
hyper-nuclear matter to the low density limit, one suspects that the
qualitative picture holds even for zero density.
Different arguments based on large $N_c$ counting, also support
the existence of the meta-stable QCD domain walls at zero chemical potential
$\mu$\cite{Forbes:2000et}.}. 

2) Making the approximation of thin flat domain walls we proceeded to
calculate the averages of quantities like
$\langle\bar{\Psi}i\sigma_{\mu\nu}\Psi\rangle$ by reducing the
interaction (\ref{eq:L4d}) to a $1+1$ dimensional system.  In this
approximation we discarded the momenta in the $x$--$y$ plane to get
(\ref{eq:decouple}).  We capture the effects of these momenta by
degeneracy factors in section~(\ref{sec:degeneracy}).  This
approximation is valid only if the physics in the $z$ direction is
independent of motion in the $x$--$y$ plane.  This approximation
breaks down when thermal (or other) fluctuations are large enough that
the physics in the $z$--$t$ directions no longer decouple from the
physics in the $x$--$y$ plane.
  
In making estimates like (\ref{eq:Spin}), another approximation we
have made is to ignore back reactions.  We have treated the domain
wall as a static background: in reality, the presence of fermions in
the domain wall would affect the structure.  What we say is that such
back reactions will not change the overall structure or scale of the
phenomenon, however, it will definitely alter the quantitative
results.  Thus, estimates like (\ref{eq:Spin}) should only be taken as
qualitative approximations to the structure in the domain wall.  A
comprehensive analysis would take into account the effects of fermions
on the domain walls through additional interactions to
(\ref{eq:Leff}).  These interactions, however, would not alter the
$U(1)$ nature of the $\eta'$ field, and thus the basic domain wall
structure would still be present.  Also, it is unlikely that the
back-reaction of the electromagnetic fields can substantially affect
the domain wall structure or the alignment of the spins.  Indeed, the
domain walls and the spin alignment are due to QCD interactions on the
scale $\Lambda_{\mathrm{QCD}}$.  Any back-reaction, would be
suppressed at least by a factor of $\alpha\sim1/137$, thus, the
quantitative results might be altered, but we expect the qualitative
behaviour and orders of magnitudes to be preserved.

3) The next step was to estimate the strengths of the generated fields
by using dimensional arguments and considering a collection of dipoles
aligned in the domain wall background arriving at the estimates
(\ref{eq:EB}).  The actual fields generated will be sensitive to the
geometry and dynamics of the domain walls: this is something that we
need to understand much better.  We have captured these effects in the
unknown scale length $\xi$ but there is much we could understand about
this.  To study these effects we will need to solve Maxwell's
equations (\ref{eq:Lem}), however, in the non-trivial geometry of the
domain walls we will probably have to simulate this.  Of particular
importance is the question: Can the larger fields be generated by the
motion of the domain walls?

The interaction of the fields with the plasma is also important, as
the electric fields may be screened. Simple estimates, however, show
that the screening timescale is at least as long as the other
timescales.
  
4) We have estimated the scaling of the magnetic turbulence
(\ref{eq:scale}) and showed that the fields generated by domain walls
at the QCD phase transition would be of astrophysical interest.  The
point we make here is that fields are generated and correlated on the
same length as the domain walls.  Thus, the domain walls provide a
mechanism for generating Hubble scale correlations.

Without a better understanding of the dynamics of the domain walls, we
cannot better estimate the field strengths, but it is possible that
the generated fields are quite large (nanogauss to microgauss scale)
even without any amplification.  It seems that the fields will still
require some amplification by galactic dynamos, but these primordial
seeds provide the large scale correlation lengths that have been
difficult to achieve through other mechanisms.

We have seen that domain walls at the QCD phase transition may provide
a nice way to resolve some of the problems with generating large scale
magnetic fields.  It is important to note that the mechanism described
here is seated in well-established physics and makes definite
predictions.  The only free ``parameter'' is a correlation length
$\xi$ which affects only the strengths of the fields generated.  This
parameter is not really free, but represents our lack of understanding
of the formation and dynamics domain walls.  As this understanding is
improved, the scale of this parameter will be fixed, and the method
will make a definite prediction about the strengths of the fields.

Thus, this mechanism is testable: it already makes a definite
prediction about the order of the field correlation lengths: If fields
are observed to have much large correlations that $500$ kpc, then
either a more efficient inverse cascade mechanism must be discovered
or another method (possibly inflationary) must be considered to
generate these correlation lengths.

\subsection{Speculations and Future Directions}
This mechanism has been a first attempt to describe astrophysical
consequences of recently conjectured QCD domain walls
\cite{Forbes:2000et}.  These walls may affect several areas of
astrophysics, both through the field generation mechanism described
here and through other effects.  We discuss several of these
applications below.

The most promising possibility to test the idea of primordial seeds is
to measure extra-galactic and extra-cluster magnetic fields.  This may
be possible in the near future through measurements of anisotropies in
the cosmic microwave background (CMB) radiation spectrum
\cite{Seshadri:2000ky,Kahniashvili:2000} or through non-thermal
radiation from Compton scattering of relativistic electrons off of the
CMB \cite{Rephaeli:2000}.  Current CMB observations
\cite{Jedamzik:1999bm} place upper bounds on the strengths of
primordial fields, which the fields described in this paper respect.
There are other limits imposed on the strengths of primordial magnetic
fields from nucleosynthesis production rates, all of which this
mechanism respect.  For a thorough review, see \cite{Grasso:2000wj}.

There are many possible consequences of primordial magnetic fields
discussed in \cite{Grasso:2000wj}, but we point out a particularly
interesting possibility that might support domain wall generated
fields.  It seems that certain types of field structures might explain
the apparent violations of the GZK cutoff \cite{Farrar:1999bw}.  It is
likely that primordial fields from domain walls may posses a planar
structure that could assist in explaining these violations.

Independent of the magnetic fields, QCD domain walls might affect
early cosmology in another way.  As we discussed in the text, baryons
are concentrated on the wall causing inhomogeneities in the baryon
density.  While it seems that QCD domain walls will decay before $T =
1$ MeV, the inhomogeneity in baryon density must be redistributed by
diffusion and other dissipative processes and thus inhomogeneities
might persist that affect nucleosynthesis.

Recent measurements of the CMB by BOOMERANG \cite{Lange:2000iq} and
MAXIMA \cite{Balbi:2000tg} lead to a value of baryon density
$\Omega_B$ is larger than the value allowed by the conventional model.
The agreement can be achieved \cite{Kurki-Suonio:2000kx} through
inhomogeneous big bang nucleosynthesis (IBBN) if regions of baryon
inhomogeneity are separated by a distance scale of about $70$ km (at
$T=1$ MeV), and if these regions have a planar structure with high
surface to volume ratio. Models have been proposed whereby such a
planar structure can occur \cite{Layek:2001rx}, however, we suggest
that QCD domain walls might automatically create this kind of
structure with the appropriate scale if the baryon diffusion and other
dissipative processes are sufficiently slow (or if something extends
the lifetime of the QCD domain walls beyond the estimate
(\ref{eq:DWtau})).  Conversely, nucleosynthesis provides another
constraint to check the validity of QCD domain wall properties.
Additional work is required \cite{Forbes:2001a} before any definite
statements can be made.

At this point, we would like to comment on a speculative application
of QCD domain walls that may be of significant astrophysical interest.
Given that QCD domain walls are very stable at high densities
\cite{Son:2000fh}, it is not inconceivable that the accumulation of
baryons along the domain wall by a mechanism similar to that described
in section~(\ref{sec:align}) may stabilize the the domain walls so
that they survive much longer than (\ref{eq:DWtau}).  If this is the
case, then QCD domain walls might form the seeds for primordial
baryonic compact objects.  Such objects would certainly affect
nucleosynthesis, the cosmic microwave background radiation spectrum
and possible structure formation.  The result, however, would be
stable, cold, ``invisible'' baryonic matter that would contribute to
the dark matter.  In addition, this matter would have QCD scale
cross-section recently suggested for dark matter \cite{Spergel:1999mh}
to explain several discrepancies with the standard cold dark matter
picture.  Besides that, this matter would offer a simple explanation
of why observations find $\Omega_{Dark}\sim \Omega_{B}$ to within an
order of magnitude. This fact is extremely difficult to explain in
models where a dark matter candidate is related to a physics
independent of baryogenesis.  Without further calculations, it is not
possible to make even qualitative predictions about these
speculations, but, as our understanding improves, such a model would
be able to make definite predictions.  Furthermore, the physics of QCD
domain walls may lend itself to signatures that can be tested in
relativistic heavy ion collisions.  This might provide some concrete
foundations for QCD phenomena affecting cosmology and astrophysics.
Likewise, the accuracy of nuclear abundance measurements and CMB
anisotropy measurements could provide serious constraints on the
behaviour of QCD phenomena related to domain walls.  Thus, while we
cannot yet seriously advocate the idea of compact primordial baryonic
objects forming from this mechanism, one should at least keep them in
mind when studying these and similar processes.

We would like to close by emphasizing the relationship that is
developing between astrophysics and particle physics at the QCD scale.
Particle physics provides concrete models for astrophysical phenomena
that, as our understanding of fundamental physics increases, has
definite predictive power.  Thus, there is the potential to ground
many astrophysical phenomena in well founded, testable physics.  In
return, astrophysical observations provide a means for testing
particle physics theories under conditions not possible on earth.  We
eagerly await the fruitful developments that will result from this
reciprocal relationship.

\begin{acknowledgments}
  This work was supported by the NSERC of Canada.  We would like to
  thank R.~Brandenberger for many useful discussions.  AZ wishes to
  thank: P.Steinhardt and M. Turner for valuable discussions;
  L.~McLerran and D.~Son for discussions on Silk damping; and
  M.~Voloshin and A. Vainshtein for discussions on the magnetic
  properties of domain walls.  MF wishes to thank F.~Wilczek and
  K.~Rajagopal for useful discussions about high density matter and
  astrophysics.
\end{acknowledgments}

\end{document}